\documentclass[11pt]{revtex4-2}

\usepackage{graphicx}
\usepackage{mathtools}
\usepackage{bm}
\usepackage{tabularray}
\UseTblrLibrary{booktabs}

\usepackage[hidelinks]{hyperref}

\usepackage{xcolor}
\definecolor{forestgreen}{rgb}{0.33,0.61,0.34}

\definecolor{newcolor}{HTML}{ff7f00}

\definecolor{oceanblue}{HTML}{79a3c9}

\definecolor{bluegreen}{HTML}{00B3B8}

\usepackage[normalem]{ulem}

\newcommand{\stkout}[1]{\ifmmode\text{\sout{\ensuremath{#1}}}\else\sout{#1}\fi}

\begin{document}

\title{Cooperation and the social brain hypothesis in primate social networks}

\author{Neil G. MacLaren$^1$}
\author{Lingqi Meng$^{1,\dagger}$}
\author{Melissa Collier$^2$}
\author{Naoki Masuda$^{1,3,4,*}$}

\affiliation{$^1$Department of Mathematics, State University of New York at Buffalo, NY 14260-2900, USA}
\affiliation{$^2$Department of Biology, Georgetown University, DC 20057, USA}
\affiliation{$^3$Institute for Artificial Intelligence and Data Science, University at Buffalo, State University of New York at Buffalo, USA}
\affiliation{$^4$Center for Computational Social Science, Kobe University, Kobe, 657-8501, Japan}
\affiliation{$^{\dagger}$Present address: Science for Life Laboratory, KTH – Royal Institute of Technology, Stockholm, SE-17121 Sweden}
\affiliation{$^{*}${\rm naokimas@buffalo.edu}}
\date{\today}



\begin{abstract}
  The social brain hypothesis posits that species with larger brains tend to have greater social complexity.  Various lines of empirical evidence have supported the social brain hypothesis, including evidence from the structure of social networks.  Cooperation is a key component of group living, particularly among primates, and theoretical research has shown that particular structures of social networks foster cooperation more easily than others.  Therefore, we hypothesized that species with a relatively large brain size tend to form social networks that better enable cooperation.  In the present study, we combine data on brain size and social networks with theory on the evolution of cooperation on networks to test this hypothesis in non-human primates.  We have found a positive effect of brain size on cooperation in social networks even after controlling for the effect of other structural properties of networks that are known to promote cooperation.
\end{abstract}

\maketitle

\section{Introduction}
\label{sec:intro}

The social brain hypothesis states that, among primates, brain size is positively associated with social complexity \cite{Dunbar1998EvolAnthropol}. Group size, in terms of the number of individuals, is one aspect of social complexity \cite{kappeler2019}. Studies have found a positive association between brain size and the typical sizes of defined social units \cite{Dunbar1992JHumanEvol, Dunbar1998EvolAnthropol} as well as more focused subgroups, such as the number of regular social contacts an individual maintains \cite{KudoDunbar2001AnimBehav, Bickart2011NatNeurosci, Kanai2011ProcRSocB, Lewis2011Neuroimage}. However, some studies have found stronger relationships between brain size and other behaviors, such as diet \cite{decasien2017}, or found the relationship between brain size and group size to be relatively weak \cite{street2017} or inconsistent across data sets \cite{powell2017}.

Although group size is the most studied potential correlate of brain size in the social brain hypothesis literature, it is not the only one \cite{dunbar2017}.
Patterns of behavior between individuals in differentiated pairwise interactions can also be thought of as an important component of social complexity \cite{dunbar2017, kappeler2019, shultz2022}. Such pairwise interactions can be represented as social networks. Network science is a common tool for studying complex systems, and researchers have investigated several network indices in relation to the social brain hypothesis. Examples include the number of connections an individual maintains (also known as the node's degree) \cite{KudoDunbar2001AnimBehav, Kanai2011ProcRSocB, Bickart2011NatNeurosci}, the number of different types of connections \cite{Bickart2011NatNeurosci}, the number of individuals in a subgroup who can connect to each other by a sequence of edges (i.e., the size of strongly connected components) \cite{KudoDunbar2001AnimBehav}, the number of connections normalized by the number of individuals (called the network density) \cite{Lehmann2009ProcRSocB}, and more sophisticated measures of network structure \cite{Lehmann2009ProcRSocB, Pasquaretta2014SciRep}. The results of all of these network-based studies are largely consistent with the social brain hypothesis.

Social networks have both benefits and costs that make them relevant to the evolution of sociality. The structure of animal social networks has been suggested to affect, for example, the speed of diffusion of information, mating behavior, predator avoidance, communication efficiency, and group movement \cite{Pinterwollman2014BehavEcol, Pasquaretta2014SciRep, Krause2015book, Brask2021JCompNetw}. On the other hand, network structure determines disease transmission potential and epidemic outcomes in populations, because a pathogen can only spread if the relevant form of contact exists between two individuals. Networks with high degree heterogeneity (i.e. high variation in the number of contacts among individuals) have increased transmission potential due to the presence of superspreaders which cause rapid, explosive outbreaks of disease in a population \cite{bansal2007}. Animal social networks that we observe today may therefore be a result of evolutionary processes in which more advantageous network structures have proliferated at the expense of less advantageous structures under restrictions imposed by the environment and trade-offs between different objectives.

One function for which social networks are particularly relevant is cooperation. Individuals of various animal species cooperate with each other, even cooperating with non-kin and in social dilemma situations in which non-cooperation is more lucrative than cooperation \cite{Maynardsmith1982book, kappeler2006, noe2006, cheney2011, croft2015, mcauliffe2015, gokcekus2021} (but see Refs.~\cite{Cluttonbrock2009Nature, cheney2010, cheney2011}, which point out that empirical evidence of cooperation in animal groups remains relatively scarce). Although cooperation under social dilemmas is an evolutionary puzzle, theoretical research has suggested various mechanisms enabling cooperation, such as direct reciprocity (i.e., repeated interaction) and indirect reciprocity (specifically, reputation-based mechanisms) \cite{Fudenberg1998book, Nowak2006Science}. Signaling, including symbolic communication, has been proposed as another mechanism that can enable cooperation \cite{smith2010}, and recent theory has suggested that structured populations may facilitate the spread of cooperation in the presence of symbolic communication when compared to well-mixed populations \cite{salahshour2020}. The structure of social networks is itself one mechanism that may promote cooperation, a concept known as network reciprocity \cite{Nowak2006Science, Szabo2007PhysRep, perc2013, perc2017, takacs2021}. Specifically, a relatively small node degree (i.e., the number of neighboring individuals per individual) \cite{Ohtsuki2006Nature, Allen2017Nature} and heterogeneity among individuals in the network in terms of the degree \cite{santos2005, santos2006} can both promote cooperation compared to well-mixed populations depending on the assumptions underlying the evolutionary process models. In addition, it has long been known that clustering of the network (i.e., abundance of short cycles such as triangles and squares) promotes cooperation, which is often referred to as spatial reciprocity \cite{Nowak2006Science, NowakMay1992Nature, hauert2001}.

The purpose of the present study is to investigate the link between the social brain hypothesis and cooperation in social networks. While cooperation occurs in various animal taxa \cite{croft2015, noe2006}, here we focus on non-human primates because both brain size and social network data are available for many primate species. Recently developed mathematical theory enables us to quantify the extent to which a network itself supports the spread of cooperation \cite{Allen2017Nature}. We use this theory and test whether species with larger brains form social networks that foster cooperation to a greater extent than networks for other species.

\begin{figure}
  \centering
  \includegraphics[width=0.99\textwidth]{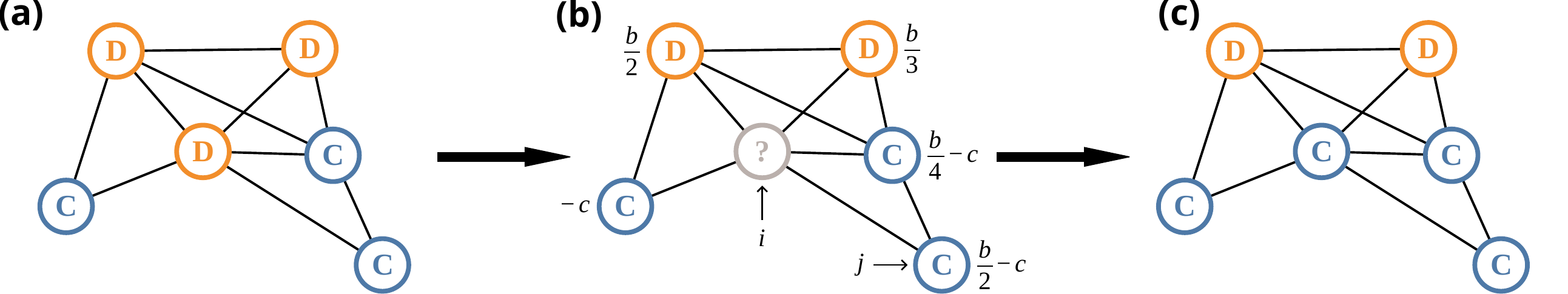}
  \caption{
    A single round of the death-birth process with selection on birth. (a) Each player gains an averaged payoff by interacting with all its neighbors. We denote cooperator and defector by C and D, respectively. (b) We select a node to be updated uniformly at random. In our example we choose the node denoted by $i$. Then, one of $i$'s neighbors, denoted by $j$, whose payoff value is shown, will replace $i$. We select as $j$ each neighbor of $i$ with probability proportional to its expected payoff; the probability to select this $j$ is given by $[1+\eta (b/2-c)] / [1+\eta (-c) + 1+\eta (b/2) + 1+\eta (b/3) + 1+\eta (b/4-c) + 1+\eta (b/2-c)] = [12+6\eta (b-2c)]/[60+\eta (19b-36c)]$, where $\eta \ll 1$ denotes the selection strength, $b$ denotes the benefit from cooperating, and $c$ denotes the cost of cooperating. (c) In this example, $j$ is a cooperator and replaces the defector on the $i$th node.
  }
  \label{fig:DBprocess}
\end{figure}

Specifically, using game theory and the properties of random walks on networks, Allen et al. \cite{Allen2017Nature} derived an expression which predicts, for an arbitrary weighted, undirected network, how much larger the benefit $b$ of cooperating must be, when compared to its cost $c$, in order to favor the spread of cooperation. The theory by Allen and colleagues relies on a death-birth process which, given an invading cooperator and assuming no mutation, leads to fixation of either cooperation or defection (Fig.~\ref{fig:DBprocess}; see the Materials and methods section for details). For a given network, cooperation fixates with a higher probability when $b/c$ is larger in general. In particular, cooperation fixates with a probability larger than a baseline probability when $b/c$ is larger than a threshold value, denoted $(b/c)^*$, and the $(b/c)^*$ value depends on the network structure (see the Materials and methods section for mathematical details). Because a small $(b/c)^*$ value implies that cooperation fixates relatively easily for a relatively small value of $b/c$, networks with a small $(b/c)^*$ value favor the spread of cooperation. Our hypothesis is that nonhuman primate species with larger neocortex ratios are associated with social networks that have lower $(b/c)^*$ values.

\section{Materials and methods}
\label{sec:methods}

\subsection{Evolutionary game dynamics and the derivation of $(b/c)^*$}

In this section, we explain the derivation of $(b/c)^*$ for any given network under the weak selection limit, following~\cite{Allen2017Nature}.

\subsubsection{Networks and discrete-time random walk}

We assume connected and undirected networks with $N$ nodes. For each pair of nodes $i, j \in \{1, \ldots, N\}$, we denote the edge weight by $w_{ij} \geq 0$. We set $w_{ij}=0$ if there is no edge $(i, j)$. We allow self-loops, i.e., the case of $w_{ii} > 0$~\cite{Allen2017Nature}. The weighted degree of node $i$, also referred to as node strength, is given by $s_i = \sum_{j=1}^N w_{ij}$. 

We start with explaining discrete-time random walks on networks because they are necessary for describing both the evolutionary game dynamics and the derivation of $(b/c)^*$.
By definition, a discrete-time random walk on the network is simple if the walker located on the $i$th node moves to any $j$th node in a single time step with probability proportional to $w_{ij}$, i.e., with probability $p_{ij} = w_{ij}/s_i$.  The transition probability matrix $P = (p_{ij})$ of the simple random walk is given by $P = D^{-1}W$, where $D = \mathrm{diag}(s_1, \ldots, s_N)$, i.e., the diagonal matrix whose diagonal entries are equal to  $s_1, s_2, \ldots, s_N$, and $W = (w_{ij})$ with $i, j \in \{1, \ldots, N\}$ is the weighted adjacency matrix. Let $\boldsymbol \pi = (\pi_1, \ldots, \pi_N)$ be the stationary probability vector of the random walk with transition probability matrix $P$. Vector $\boldsymbol \pi$ is the solution of $\boldsymbol \pi P = \boldsymbol \pi$ satisfying $\sum_{i=1}^N \pi_i = N$. It holds true for undirected networks that \cite{aldous1995, masuda2017}
\begin{equation}
\pi_i = \frac{s_i}{\sum_{\ell=1}^N s_\ell},\quad i\in \{1, \ldots, N\}.
\label{eq:pi_i}
\end{equation}

\subsubsection{Gift-giving game and evolutionary dynamics under the death-birth updating rule}

We use the gift-giving game, also called the donation game, which is a subtype of the prisoner's dilemma game. In the gift-giving game, which is a two-player game, one player, called the donor, decides whether or not to pay a cost $c$ $(>0)$. If the donor pays $c$, which we refer to as cooperation, then the other player, called the recipient, receives benefit $b$, which we assume to be larger than $c$. If the donor decides not to pay $c$, which we refer to as defection, then the donor does not lose anything, and the recipient does not gain anything. We assume that each player plays the game with each neighbor once as donor and another time as recipient in a single round of evolutionary dynamics. Then, the payoff matrix of the gift-giving game between a pair of players is given by
\begin{equation}
\bordermatrix{
& \mathrm{C} & \mathrm{D}\cr
\mathrm{C} & b-c & -c\cr
\mathrm{D} & b & 0 },
\end{equation}
where C and D represent cooperation and defection, respectively, and the payoff values represent those for the row player. 

We set $x_i=0$ or $x_i=1$ when the $i$th player is defector or cooperator, respectively. Then, the state of the entire network is specified by a binary vector $\boldsymbol x =(x_1, \ldots, x_N) \in \{0, 1\}^N$. The payoff of the $i$th node averaged over all its neighbors is given by
\begin{align}
    f_i(\boldsymbol x) = -c x_i + b \sum_{j=1}^N p_{ij}x_j.
\end{align}
We set the reproductive rate of node $i$ in state $\boldsymbol x$ by
\begin{align}
    R_i (\boldsymbol x) = 1 + \eta f_i(\boldsymbol x),
\end{align}
where $\eta~(\ge 0)$ represents the strength of natural selection. When $\eta \to 0$, the payoff, $f_i(\boldsymbol x)$ only weakly impacts the selection, which is called the weak selection regime. A justification of weak selection is that, in reality, many different factors may contribute to the overall fitness of an individual, and the prisoner's dilemma game may be just one such contributor~\cite{Ohtsuki2006Nature, Allen2017Nature}.

We assume evolutionary dynamics of the gift-giving game driven by the death-birth process with selection on birth~\cite{Ohtsuki2006Nature, Allen2017Nature}. By definition, we first select a node to be updated (i.e., die), denoted by $i$, uniformly at random. Second, we select one of the neighbors of the $i$th node, denoted by $j$, for reproduction (i.e., give birth), with the probability proportional to $w_{ij} R_j(\boldsymbol x)$. Third, $i$ copies the type (i.e., defection or cooperation) of $j$.
These three steps constitute a single round of the evolutionary dynamics; see Fig.~\ref{fig:DBprocess} for a schematic. 

\subsubsection{Fixation probability for cooperation and the expression of $(b/c)^*$}

Because we omitted mutation, the death-birth process in any finite network eventually terminates in the state in which all individuals are uniformly cooperators or defectors. We call these final states fixation of cooperation or defection. According to a standard convention, we assume that the initial state contains one cooperator node and $N-1$ defector nodes and that each node is the unique initial cooperator with the equal probability $1/N$. We denote by $\rho_{\mathrm{C}}$ the probability that cooperation fixates. Defection fixates with probability $1 - \rho_{\mathrm{C}}$. We say that natural selection favors cooperation if $\rho_{\mathrm{C}} > 1/N$~\cite{Nowak2004Nature, Ohtsuki2006Nature, Nowak2006book, Allen2017Nature}. 

Allen et al. showed that~\cite{Allen2017Nature} 
\begin{align}
    \rho_{\mathrm{C}} = \frac{1}{N} + \frac{\eta}{2N}\left[-c\tau_2 + b(\tau_3 - \tau_1)\right] + O(\eta^2),
    \label{rho_c}
\end{align}
where
\begin{align}
    \tau_k = \sum_{i=1}^N\sum_{j=1}^N \pi_i p_{ij}^{(k)}t_{ij},
    \label{tau k}
\end{align}
$p_{ij}^{(k)}$ is the $(i,j)$th entry of matrix $P^k$, which implies that $p_{ij}^{(1)} = p_{ij}$, and 
\begin{align}
    t_{ij} =
    \begin{cases} 
      0 & \mathrm{if} \ i=j, \\
      1 + \frac{1}{2}\sum_{k=1}^N(p_{ik}t_{jk} + p_{jk}t_{ik}) & \mathrm{otherwise}.
   \end{cases}
   \label{t_ij}
\end{align}
Equation (\ref{t_ij}) implies that $t_{ij} = t_{ji}$ is the mean coalescence time of two random walkers when one walker is initially located at the $i$th node and the other walker is initially located at the $j$th node. Note that $p_{ij}^{(k)}$ is the $k$-step transition probability of the random walk from the $i$th to the $j$th node. Therefore, $\tau_k$ is the expected value of $t_{ij}$ when $j$ is the node at which the random walker arrives after $k$ steps starting at the $i$th node under the stationary distribution \cite{Allen2017Nature}. Equation~(\ref{rho_c}) implies that the natural selection favors cooperation (i.e., $ \rho_{\mathrm{C}} > 1/N$) under weak selection if and only if
\begin{align}
    \left(\frac{b}{c} \right) > \left(\frac{b}{c}\right)^* \equiv \frac{\tau_2}{\tau_3 - \tau_1}.
    \label{bcthreshold}
\end{align}
It should be noted that the right-hand side of Eq.~\eqref{bcthreshold} only depends on the adjacency matrix of the network, $W$. Therefore, the network structure determines whether and how much natural selection favors cooperation in the present model. Note that $(b/c)^*$ is a threshold value: cooperation is predicted to fixate with a probability larger than $1/N$ when the ratio of benefit $b$ to cost $c$ of a particular cooperative behavior is larger than $(b/c)^*$. Thus, cooperation spreads more easily on networks with lower $(b/c)^*$. 

We calculated $(b/c)^*$ for each network using our in-house code in Python 3.10, which implements the procedures described in \cite{Allen2017Nature}; the code is available at \href{https://github.com/ngmaclaren/cooperation-threshold}{https://github.com/ngmaclaren/cooperation-threshold}.

\subsection{Data}
\label{sub:data}

The data for this study come from the Animal Social Network Repository (ASNR) \cite{sah2019, collier2021}. The ASNR contains 770 non-human social networks from eight animal classes and 69 species. 
For each network in this data set, nodes represent an individual animal. Edges represent a specific type of contact between two animals, such as grooming in primates and trophallaxis in ants, as well as more general contact such as group membership and spatial proximity.

There are 114 non-human primate social networks in the ASNR, including 60 grooming networks, 31 spatial proximity networks, 10 mating networks, and 13 networks with other contact types. Most sampled populations are free-ranging (84), with some captive (18) and some semi-freeranging (7) populations, as well as five populations for which the type was not recorded. There are 99 catarrhine primate networks, 13 platyrrhine networks, and 2 strepsirrhine networks. Sampling of the different contrasts represented in the ASNR is thus somewhat unbalanced but reflects the sampling effort present in the literature. 

To test our hypothesis we require that, to the best extent possible, the edges represent prosocial contacts between individuals. Other contact types, such as dominance or mating, may reflect motives that are not relevant to the spread of cooperative behaviors, and proximity-based networks may reflect individuals who are co-located by chance or interest in a common resource rather than for social interaction. We therefore used the ASNR networks with the interaction types labeled  ``grooming'', ``physical contact'', and ``overall mix''; the ``overall mix'' category captures one additional network that recorded grooming behavior. We thus obtained 67 possible networks, which we regarded as undirected weighted networks.

Thirteen out of the 67 networks yielded negative $(b/c)^*$ values, which imply that spiteful behavior evolves instead of cooperation \cite{Allen2017Nature, Su2022SciAdv}. We discarded these networks because we are interested in cooperation under social dilemma situations, and because the qualitatively different interpretation of a unit change for  $(b/c)^*$ values above and below zero (i.e., a unit change in the positive direction below zero means that spite evolves more easily, whereas a similar change above zero means that cooperation evolves less easily) violates regression modeling assumptions. Additionally, we discarded one network that was composed of two disconnected dyads and used the remaining 53 connected networks for our analysis. Most species had a single network in the repository. The exceptions were {\em Papio cynocephalus} (which had 23 networks), {\em Macaca fascicularis} (2), {\em M. fuscata} (4), {\em M. mulatta} (9), and {\em M. radiata} (2). For these species we took the median for $(b/c)^*$ and for the network-based explanatory variables explained in Section~\ref{sub:analysis} to prevent a few species, such as {\em P. cynocephalus} and {\em M. mulatta}, from dominating the set of networks to be analyzed.
In this manner, we reduced the 53 networks to observations on 17 species for further analysis (Table \ref{tab:data}).

\begin{table}
  {\renewcommand\baselinestretch{1}\selectfont
    \begin{tblr}{Q[l,f]Q[c,f]Q[c,f]Q[c,f]Q[c,f]Q[c,f]Q[c,f]Q[c,f]Q[c,f]}
      \toprule
      Species & $(b/c)^*$ & Neocortex Ratio & Brain mass & $N$ & $\langle k \rangle$ & $\langle s \rangle$ & $C$ &  $\tilde{C}_{\text{w}}$\\
      \midrule
      Sapajus apella & 12.59 & 2.25 & 66.63 & 12 & 7.17 & 7.17 & 0.69 & 0.08\\
      Macaca arctoides & 18.63 & 2.43 & 100.7 & 20 & 10.62 & 17.13 & 0.62 & 0.08\\
      Cercopithecus campbelli & 34.46 & 2.21 & 57.39 & 15 & 7.87 & 7.87 & 0.66 & 0.05\\
      Papio cynocephalus & 4.30 & 2.68 & 163.19 & 11 & 2.56 & 3.56 & 0.16 & 0.07\\
      Macaca fascicularis & 2.18 & 2.6 & 63.98 & 10.5 & 3.86 & 5.89 & 0.35 & 0.02\\
      Macaca fuscata & 12.98 & 2.45 & 102.92 & 9 & 6.52 & 92.53 & 0.91 & 0.06\\
      Ateles geoffroyi & 10.02 & 2.35 & 105.09 & 15 & 6 & 6 & 0.53 & 0.09\\
      Colobus guereza & 8.76 & 2.32 & 74.39 & 8 & 4.5 & 4.5 & 0.59 & 0.10\\
      Ateles hybridus & 11.81 & 2.35 & 103.05 & 17 & 8.47 & 794.47 & 0.81 & 0.09\\
      Macaca mulatta & 8.10 & 2.6 & 88.98 & 78 & 14.3 & 41.33 & 0.29 & 0.02\\
      Pan paniscus & 5.04 & 3.22 & 341.29 & 19 & 5.79 & 5.79 & 0.46 & 0.06\\
      Papio papio & 4.88 & 2.76 & 163.19 & 25 & 7.76 & 7.76 & 0.41 & 0.03\\
      Erythrocebus patas & 8.13 & 2.96 & 97.73 & 19 & 5.16 & 5.16 & 0.56 & 0.07\\
      Macaca radiata & 28.22 & 2.28 & 74.87 & 18 & 9.86 & 15.74 & 0.70 & 0.11\\
      Macaca sylvanus & 3.08 & 2.37 & 93.2 & 8 & 7 & 26.97 & 1 & 0.02\\
      Macaca tonkeana & 24.12 & 2.6 & 93.7 & 25 & 14.48 & 14.48 & 0.62 & 0.07\\
      Pan troglodytes & 11.42 & 3.22 & 368.35 & 24 & 8.58 & 8.58 & 0.65 & 0.08\\
          \bottomrule
    \end{tblr}
  }
  \caption{
    Properties of primate social networks returned by our selection procedures, sorted by $(b/c)^*$. 
    Values are medians of all the networks for {\em Papio cynocephalus}, {\em Macaca fasicularis}, {\em M. fuscata}, {\em M. mulatta}, and {\em M. radiata}.
    NCR: neocortex ratio, $N$: number of nodes, $\langle k \rangle$: average node degree, $\langle s \rangle$: average node strength, $C$: clustering coefficient, $\tilde{C}_{\text{w}}$: weighted clustering coefficient.
  }
  \label{tab:data}
\end{table}

We used the species-level neocortex ratio (NCR) estimate from \cite{KudoDunbar2001AnimBehav} for all but one species, {\em Colobus guereza}; a species-level NCR estimate was not available in \cite{KudoDunbar2001AnimBehav}, so we used the genus-level NCR estimate from \cite{Dunbar1992JHumanEvol}. Additionally, we used the brain mass data from \cite{smaers2021} for all species except {\em Papio papio}, for which the data is not present. For {\em Papio papio}, we used the data of the closely related species {\em P. cynocephalus} \cite{newman2004}. Because the size of several regions of the brain may correlate with social complexity \cite{Bickart2011NatNeurosci, Kanai2011ProcRSocB, Lewis2011Neuroimage}, we included overall brain mass as a relatively simple measure, when compared to the NCR, of species' neurological complexity \cite{decasien2017} that may also correlate with sociality \cite{smaers2019}. These two measures (i.e., brain mass and NCR) are highly correlated with each other (see Section~\ref{sec:results}). Given the unbalanced sampling mentioned above, we did not include controls for phylogeny, social system, foraging behavior, whether the group was free-ranging or captive, or type of behavior captured by the network. See Section \ref{sec:discussion} for further discussion of this limitation.

\subsection{Analysis}
\label{sub:analysis}

Data analysis was conducted in R \cite{R}; the code is available at \href{https://github.com/ngmaclaren/cooperation-threshold}{https://github.com/ngmaclaren/cooperation-threshold}. We used the ``MuMIn'' package \cite{barton2022} to implement the model selection procedure described below.

We fitted generalized linear models (GLMs) to test whether NCR and other variables were associated with the difficulty of cooperation, $(b/c)^*$, which we used as the dependent variable. We considered seven explanatory variables: NCR, brain mass in grams, and five network indices. The five network indices are the number of nodes in the network, denoted by $N$, the average degree over the $N$ nodes, $\langle k \rangle$, the average node strength (i.e., the average of the weighted degree over the $N$ nodes), $\langle s \rangle$, the clustering coefficient, $C$, and the weighted clustering coefficient, $\tilde{C}_{\text{w}}$. The clustering coefficient is the average over all nodes of the local clustering coefficient; the local clustering coefficient for the $i$th node is the number of triangles (i.e., ($i$, $i'$), ($i$, $i''$), and ($i'$, $i''$) are edges of the network) divided by the number of possible triangles involving the $i$th node (i.e., $k_i (k_i-1)/2$, where $k_i$ is the node $i$'s degree) \cite{wasserman1994book, newman2018book}.
The weighted clustering coefficient is calculated similarly to the unweighted version except that it uses the geometric mean of the edge weights instead of a count of edges \cite{fagiolo2007}. We include these network indices because each of these indices can affect $(b/c)^*$ regardless of the potential relationship between brain size and $(b/c)^*$ \cite{Allen2017Nature}.
Because brain mass, body mass, $\langle s\rangle$, and $\tilde{C}_{\text{w}}$ are positive and obey right-skewed distributions, we used the natural logarithm transform of each of these variables.

We began our modeling process from a position of relative ignorance, including these seven explanatory variables as predictors. By design, our outcome variable, $(b/c)^*$, is positive and continuous, suggesting a model with gamma-distributed errors. To test our choice, we built five different models, each with all seven explanatory variables, with different error models and link functions (i.e., gamma and Gaussian distributions with both inverse and log links, and a quasi-Poisson model) and calculated the deviances of each \cite{faraway2016}. As expected, the gamma models fit well ($\chi^2$ test with $d.f. = 8$; $p$ = 0.968 and 0.991 for the inverse and log links, respectively), whereas the other models did not ($p \leq 0.001$ for each). The residual deviances associated with both gamma-based models are small (inverse link: 2.90, log link: 2.01) relative to the remaining models (quasi-Poisson: 27.61, Gaussian inverse link: 236.09, Gaussian log link: 374.50), further suggesting good fit \cite{faraway2016}. Because the model with gamma-distributed errors and the log link had the minimum residual deviance, we used that model for further analysis \cite{faraway2016}.

The number of explanatory variables (i.e., seven) is relatively large given the number of observations (i.e., 17). Therefore, we ran an AIC-based model selection, as follows. First, we evaluated all possible models---excluding any model with both brain mass and NCR as predictor variables---and calculated the AICc for each model. AICc is a modification of the Akaike Information Criterion (AIC) that is preferred for model selection when data sets are relatively small \cite{burnham2002}. Specifically, AICc is defined as AIC$+ (2k^2 + 2k)/(n - k - 1)$, where $k$ is the degrees of freedom of the model and $n$ is the number of observations. When $n$ is small, AICc values increase more with each additional model parameter than the traditional AIC does; the difference between the two metrics becomes small when $n$ is large. Model selection based on AICc thus tends to support fewer model parameters at small $n$ than traditional AIC. A recommended rule is to use AICc when $n/k < 40$ \cite{burnham2002}; for a model in this study with three predictor variables, an intercept, and an error parameter, we obtain $n/k = 17/5 = 3.4 \ll 40$. We sorted all evaluated models by AICc: the models with minimal AICc values realize the best fit to the data with the fewest variables. 

\section{Results}
\label{sec:results}

\begin{figure}
  \centering
  \includegraphics[width = .6\textwidth]{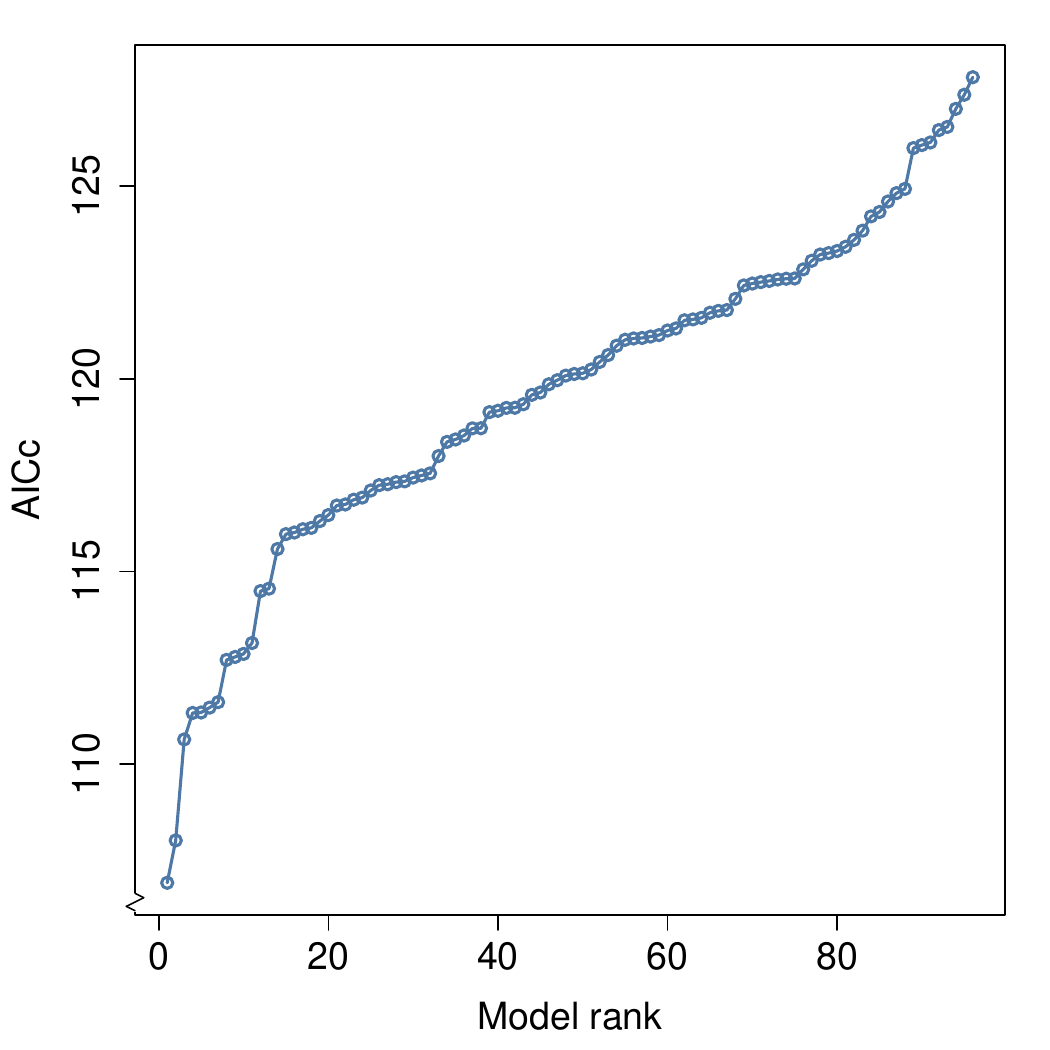}
  \caption{Akaike Information Criterion adjusted for small samples (AICc) for all possible models. Possible models are generalized linear models with gamma-distributed errors, a natural logarithm link function, and zero or more of the following explanatory variables: neocortex ratio, brain size, number of nodes, average node degree, average node strength, clustering coefficient, and weighted clustering coefficient.}
  \label{fig:AICcomp}
\end{figure}

We show the sorted AICc values for all 96 models which met our initial selection criteria in Fig.~\ref{fig:AICcomp}. Figure~\ref{fig:AICcomp} shows that the best and second-best models are fairly similar in terms of AICc, but the third-best model has somewhat poorer AICc. Setting a cutoff at $\Delta$AICc $= 3$, where $\Delta$AICc means the absolute difference in the AICc value relative to the smallest value, allows us to focus on two models that are similar in terms of AICc but clearly better than any other alternatives. We summarize these two models in Table \ref{tab:toptwo}. These two models are superior to the full model in terms of AICc (full model AICc: 132.39, Model 1: 106.91, Model 2: 108.01) and collinearity (maximum variance inflation factor for the full model: 5.08, Model 1: 1.03, Model 2: 1.02) without a substantial reduction in variance explained (full model McFadden's pseudo-$R^2$: 0.77, Model 1: 0.73, Model 2: 0.72).

\begin{table}
  {\renewcommand\baselinestretch{1}\selectfont
    \begin{tblr}{Q[l,f]Q[c,f]Q[c,f]Q[c,f]}
      \toprule
      & & & \SetCell[c=2]{c} 95\% CI\\
      Model 1 & Estimate & SE & Lower & Upper\\
      \midrule
      Intercept & 6.398 & 1.503 & 3.469 & 9.277\\
      Brain mass & $-$0.522 & 0.256 & $-0.958$ & $-0.051$\\
      $\langle k \rangle$ & 0.138 & 0.041 & 0.050 & 0.230\\
      $\tilde{C}_{\text{w}}$ & 0.944 & 0.243 & 0.390 & 1.460\\
      AICc & 106.910 &  &  & \\
      Deviance & 2.353 &  &  & \\
      Pseudo-$R^2$ & 0.73 & & & \\
      \midrule
      & & & \SetCell[c=2]{c} 95\% CI\\
      Model 2 & Estimate & SE & Lower & Upper\\
      \midrule
      Intercept & 5.809 & 1.329 & 3.134 & 8.479\\
      Neocortex ratio & $-0.827$ & 0.437 & $-1.590$ & $-0.010$\\
      $\langle k \rangle$ & 0.135 & 0.042 & 0.048 & 0.228\\
      $\tilde{C}_{\text{w}}$ & 0.844 & 0.249 & 0.281 & 1.361\\
      AICc & 108.010 &  &  & \\
      Deviance & 2.507 &  &  & \\
      Pseudo-$R^2$ & 0.72 & & & \\
      \bottomrule
    \end{tblr}
  }

  \caption{The best two models, i.e., the models with $\Delta$AICc $< 3$. The dependent variable is $(b/c)^*$. All models are generalized linear models with gamma-distributed errors and a natural logarithm link. A negative coefficient indicates that a larger value of the predictor is associated with a smaller value of $(b/c)^*$, suggesting that cooperation spreads more easily on a network. SE stands for the standard error; CI stands for confidence interval; $\langle k \rangle$ and $\tilde{C}_{\text{w}}$ represent average degree and the weighted clustering coefficient, respectively.}
  \label{tab:toptwo}
\end{table}

\begin{figure}
  \centering
  \includegraphics[width = .6\textwidth]{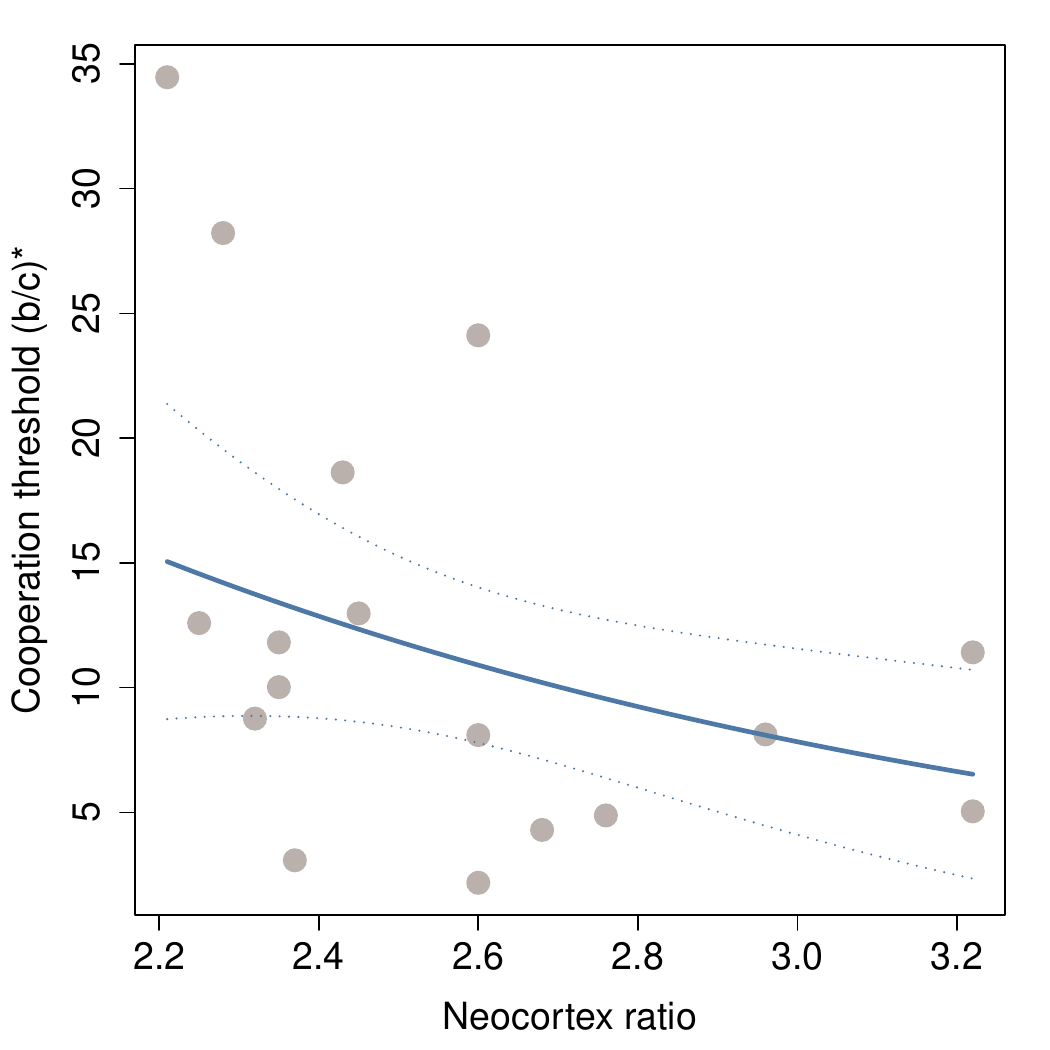}
  \caption{Threshold for cooperation, $(b/c)^*$, as a function of the neocortex ratio. Each circle represents a primate species. The solid line represents the predicted $(b/c)^*$ given median values for average degree and weighted clustering coefficient. The dotted lines indicate twice the standard error of prediction.}
  \label{fig:modelfig}
\end{figure}

The two best models both include a measure of brain size---overall brain mass in Model 1 and NCR in Model 2---and two network features: average node degree $\langle k \rangle$ and the weighted clustering coefficient $\tilde{C}_{\text{w}}$ (Table~\ref{tab:toptwo}). As is expected given the correlation between brain mass and NCR in this data ($r = 0.843$, mentioned above), coefficient estimates for the two models are similar: the coefficients on both brain size variables are both negative, whereas the coefficients on average node degree and weighted clustering are positive. Thus, we find that, when average degree and weighted clustering are held constant, brain size is inversely associated with $(b/c)^*$ in this data: primates with larger brains are associated with social networks that favor the spread of cooperation. We visualize the association between NCR and $(b/c)^*$, controlling for average degree and weighted clustering, in Fig.~\ref{fig:modelfig}.

\begin{figure}
  \centering
  \includegraphics[width = .99\textwidth]{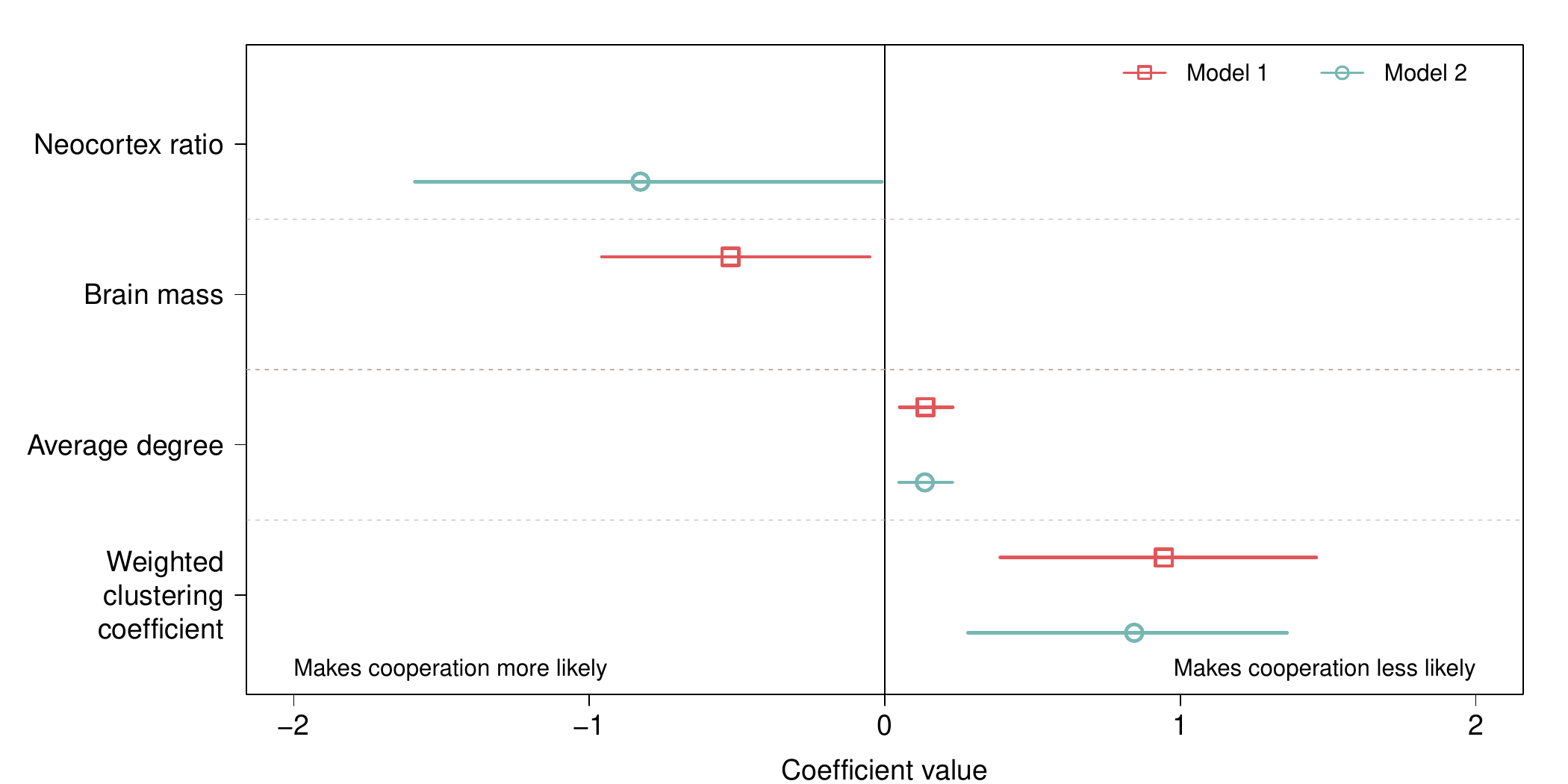}
  \caption{Coefficient estimates for the two models with $\Delta$AICc $< 3$. Both models are generalized linear models with gamma-distributed errors and a natural logarithm link function and express the association beteween three explanatory variables and $(b/c)^*$. Each model is represented by a different color and type of marker. The markers represent the coefficient values. The lines represent the profile likelihood 95\% confidence intervals.
}
  \label{fig:searchfig}
\end{figure}

Although overall trends in the data support the social brain hypothesis, there is substantial uncertainty in the coefficient estimates (see Table \ref{tab:toptwo}). We visualize this uncertainty in Fig.~\ref{fig:searchfig}, which shows the point estimate for each coefficient (open markers) in both models (indicated by color and marker shape) along with the profile likelihood 95\% confidence intervals (horizontal lines). The confidence intervals are all relatively wide compared with the magnitude of the coefficient, suggesting that the size and noisiness of our data inhibit our ability to make precise estimates of the relationship between brain size and $(b/c)^*$. 

Finally, we note that neither of the best two models has more than three explanatory variables, suggesting that adding more explanatory variables would not be useful in better explaining $(b/c)^*$ across the different networks. This observation indicates that our data do not support differentiating between the effects of brain mass and neocortex ratio by, for example, including one as a control on the other in a regression model. The failure of group size $N$ to appear in the best models is also notable, which we discuss in the Discussion section. This finding suggests that the relationship between brain size and the spread of cooperative behavior may be independent of group size. Finally, neither the average weighted degree $\langle s \rangle$ nor the unweighted clustering coefficient $C$ appeared in the five best models.

\section{Discussion}
\label{sec:discussion}

Our findings suggest that primate species with larger brains tend to form networks which, based on results from game theory \cite{Allen2017Nature}, support the spread of cooperative behaviors.
Thus, our primary results are consistent with the social brain hypothesis. Our results are also consistent with previous findings on the effect of network structure on cooperation in primates \cite{Voelkl2009BiolLett}.

Group size and NCR are only weakly correlated with each other in our data ($r=0.203$). This result is only marginally consistent with previous studies, which showed a strong association between group size, which has been used as a proxy for social complexity \cite{dunbar2017}, and NCR; this association is a central result supporting the social brain hypothesis \cite{dunbar2021}. A weak association between group size and NCR in our data may be due to different definitions of the group size used in our study and the previous ones. The group size used in this study is the observed number of individuals in a single group. That group was captive in some studies, and there may be other constraints on the observed group size in a particular study that may make the group size value different from what may be typical in wild populations. This difference may have depressed the relationship between group size and the NCR, and also the relationship between group size and cooperation in the present study. Alternatively, we note that primate groups may form for a variety of reasons, such as protection from predators, which neither relate specifically to cooperation nor necessarily indicate an increased cognitive demand on group members. By including group size as a potential predictor of the cooperation threshold, we hypothesized that groups with larger size will have consistent differences in their cooperation threshold from groups of smaller size. We found that this is not the case within the limits of our analysis. Thus, our present findings are orthogonal to previous tests of the social brain hypothesis.

Our data indicate a notable level of uncertainty in the observed trends. An important reason for this uncertainty is the relatively small sample size of our final data set. Additionally, as we described in Section \ref{sec:methods}, our models do not control for a variety of factors, such as phylogeny and study design. Omitting these variables might account for some of our reported error variance. However, these factors are represented in a very uneven way in the primate networks available in the ASNR, limiting our ability to control for them in our models in a meaningful way. For example, of the 17 species in our final data set, 11 are cercopithecine primates and of those, seven are macaques. No lemur or other strepsirrhine species are represented at all. The situation is even more extreme in the data set prior to aggregation to the species level, in which 44 of the 53 networks came from two cercopithecine genera: {\em Papio} (23) and {\em Macaca} (21). Thus, although the phylogenetic signal in group size and related variables has been previously shown to be relatively weak \cite{kamilar2013}, we are limited in our ability to control for phylogenetic effects that may be present in our model. We face a similar situation in attempting to control for study design, which can affect the structure of observed networks \cite{collier2022}: in our data, most groups were sampled according to social group membership, none were sampled according to a geographic area, and the only captive groups were 10 of the 21 macaque networks. Finally, some uncertainty in our estimates may be due to the brain mass and NCR measurements themselves, which are difficult to obtain and thus not based on large samples of individuals nor available for all species \cite{stephan1981}. Because of these conditions, we have chosen to simplify our model to accommodate a small sample \cite{matuschek2017}, rather than take a maximal approach, with which we would include as many theoretically important variables in the model as possible \cite{barr2013, mcelreath2016}. We recognize that our decisions reduce both the sample size and the potential generalizability of our study \cite{yarkoni2022}. Additional data from species more evenly spread across primate taxa will help address these concerns, as well as support mediation analysis to better test between competing causal hypotheses.

From a theoretical perspective, our work is also limited by the assumptions made by Allen et al.'s theory \cite{Allen2017Nature}. Specifically, their theory assumes fixed, undirected networks and binary strategies (i.e., cooperation or defection). Such assumptions do not realistically represent primate social networks, which may be dynamic, have asymmetric ties (i.e., individual A grooms individual B more than the reverse), and be characterized by complex behavioral strategies. This lack of an explicit connection between models and reality has been recognized as a major challenge in evolutionary game theory \cite{jusup2022}.
The strength of Allen et al.'s results and others is in providing insight into the general mechanisms of the evolution of cooperation \cite{akcay2020}. Our study addresses this gap by showing that predictions from the social brain hypothesis, based on observations, are in line with those from evolutionary game theory.

Despite these caveats, the present results allow us to make several additional observations. For example, we observe that cooperation spreads less easily on networks in which individuals tend to have many social partners (i.e., large average degree) or tend to form clusters (i.e., connected triangles). While the former observation agrees with the literature \cite{Ohtsuki2006Nature, Allen2017Nature}, the latter is apparently inconsistent with the concept of spatial reciprocity, which states that high clustering in networks promotes cooperation \cite{NowakMay1992Nature, hauert2001, Nowak2006Science}. In fact, results supporting spatial reciprocity have been derived for the fraction of cooperators in the quasi-stationary state of evolutionary dynamics in relatively large networks rather than the fixation probability for the cooperator strategy; we examined the latter quantity in this study. The effect of clustering on the fixation probability for cooperation is not systematically known. For example, some numerical simulations suggest that clustering, which is present in most empirical networks, does not facilitate the fixation of cooperation \cite{Ohtsuki2006Nature, wu2014}. Therefore, our results are in fact not contradictory to the known results for spatial reciprocity, and fixation of cooperation in clustered networks remains to be investigated.

Cooperative group living is often advantageous in the animal kingdom because it can provide protection from predators and increase the efficiency of foraging tactics \cite{alexander1974, kappeler2006}.
However, one of the most commonly cited disadvantages to cooperative group living is the increase in disease transmission potential \cite{alexander1974, freeland1976}. In fact, previous work suggests that the average degree is the most important aspect of network structure in determining the transmission potential for pathogens on a network \cite{collier2022}. Our results show that average degree is negatively associated with the evolution of cooperation, a finding supported by previous theoretical work \cite{Ohtsuki2006Nature}. Given that small average degrees are beneficial for both enhancing cooperation and reducing pathogen transmission opportunity, cooperation and protection against disease transmission potential might have coevolved through a decrease in the average degree of social networks. Maintaining contacts is also costly for individuals. However, a large average degree helps robustness of networks against node and edge failures \cite{cohen2010, Brask2021JCompNetw}. We may be able to further discussion of the evolution of network structure and social brain hypotheses by simultaneously taking into account multiple functions of animal society such as cooperation, protection against infection, robustness, and communication efficiency.

The present work also opens avenues for further work to explore the intersection between the social brain hypothesis, networks, and cooperation. For example, most of the social networks in our sample are grooming networks. However, network structure may vary according to the type of prosocial contact even for the same species of animals \cite{collier2022}. It is not currently known if differences in network structure associated with different behaviors also reflect differences in the spread of cooperation or other indices of social complexity. Furthermore, the spread of spite on ostensibly prosocial networks is an important possibility, but insufficiently characterized. Although further comparative work along these lines is currently limited by available data \cite{sah2019}, various technological and algorithmic developments of automatic data collection \cite{Brask2021JCompNetw, krause2013} are expected to allow us to access more data and explore these topics in the near future.

\section*{Acknowledgments}
 
N. Masuda acknowledges support from AFOSR European Office (under Grant No. FA9550-19-1-7024), the Japan Science and Technology Agency (JST) Moonshot R\&D (under Grant No. JPMJMS2021), the National Science Foundation (under Grant Nos. 2052720 and 2204936), and JSPS KAKENHI (under Grant Nos. JP 21H04595 and 23H03414). M. Collier acknowledges funding from the Morris Animal Foundation (under Grant No. D22ZO-059). We thank Shweta Bansal for her thoughtful comments on this work. We thank Pratha Sah, Jose Mendez, Grant Rosensteel, Elly Meng, and Sania Ali for their contributions to the development and growth of the Animal Social Network Repository.

\section*{Author contributions}

N.M. designed the study. L.M. and N.G.M. analyzed the data, supervised by N.M. M.C. curated the data. N.M. and N.G.M. wrote the manuscript with contributions by L.M. and M.C. All authors reviewed and approved the manuscript.

\section*{Data Availability}

The data for this study comes from the Animal Social Network Repository, Version 2, located at \url{https://github.com/bansallab/asnr} (DOI: 10.5281/zenodo.7595404) \cite{collier2021, sah2019}.

\section*{Competing interests}

The authors declare no competing interests.

\bibliographystyle{unsrtnat}
\bibliography{refs.bib}

\begin{thebibliography}{74}
\providecommand{\natexlab}[1]{#1}
\providecommand{\url}[1]{\texttt{#1}}
\expandafter\ifx\csname urlstyle\endcsname\relax
  \providecommand{\doi}[1]{doi: #1}\else
  \providecommand{\doi}{doi: \begingroup \urlstyle{rm}\Url}\fi

\bibitem[Dunbar(1998)]{Dunbar1998EvolAnthropol}
R.~I.~M. Dunbar.
\newblock The social brain hypothesis.
\newblock \emph{Evolutionary Anthropology: Issues, News, and Reviews},
  6\penalty0 (5):\penalty0 178--190, 1998.

\bibitem[Kappeler et~al.(2019)Kappeler, Clutton-Brock, Shultz, and
  Lukas]{kappeler2019}
P.~M. Kappeler, T.~Clutton-Brock, S.~Shultz, and D.~Lukas.
\newblock Social complexity: patterns, processes, and evolution.
\newblock \emph{Behavioral Ecology and Sociobiology}, 73:\penalty0 5, 2019.

\bibitem[Dunbar(1992)]{Dunbar1992JHumanEvol}
R.~I.~M. Dunbar.
\newblock Neocortex size as a constraint on group size in primates.
\newblock \emph{Journal of Human Evolution}, 22\penalty0 (6):\penalty0
  469--493, 1992.

\bibitem[Kudo and Dunbar(2001)]{KudoDunbar2001AnimBehav}
H.~Kudo and R.~I.~M. Dunbar.
\newblock Neocortex size and social network size in primates.
\newblock \emph{Animal Behaviour}, 62\penalty0 (4):\penalty0 711--722, 2001.

\bibitem[Bickart et~al.(2011)Bickart, Wright, Dautoff, Dickerson, and
  Barrett]{Bickart2011NatNeurosci}
K.~C. Bickart, C.~I. Wright, R.~J. Dautoff, B.~C. Dickerson, and L.~F. Barrett.
\newblock Amygdala volume and social network size in humans.
\newblock \emph{Nature Neuroscience}, 14\penalty0 (2):\penalty0 163--164, 2011.

\bibitem[Kanai et~al.(2012)Kanai, Bahrami, Roylance, and
  Rees]{Kanai2011ProcRSocB}
R.~Kanai, B.~Bahrami, R.~Roylance, and G.~Rees.
\newblock Online social network size is reflected in human brain structure.
\newblock \emph{Proceedings of the Royal Society B: Biological Sciences},
  279\penalty0 (1732):\penalty0 1327--1334, 2012.

\bibitem[Lewis et~al.(2011)Lewis, Rezaie, Brown, Roberts, and
  Dunbar]{Lewis2011Neuroimage}
P.~A. Lewis, R.~Rezaie, R.~Brown, N.~Roberts, and R.~I.~M. Dunbar.
\newblock Ventromedial prefrontal volume predicts understanding of others and
  social network size.
\newblock \emph{NeuroImage}, 57\penalty0 (4):\penalty0 1624--1629, 2011.

\bibitem[DeCasien et~al.(2017)DeCasien, Williams, and Higham]{decasien2017}
A.~R. DeCasien, S.~A. Williams, and J.~P. Higham.
\newblock Primate brain size is predicted by diet but not sociality.
\newblock \emph{Nature Ecology \& Evolution}, 1\penalty0 (5):\penalty0 0112,
  2017.

\bibitem[Street et~al.(2017)Street, Navarrete, Reader, and Laland]{street2017}
S.~E. Street, A.~F. Navarrete, S.~M. Reader, and K.~N. Laland.
\newblock Coevolution of cultural intelligence, extended life history,
  sociality, and brain size in primates.
\newblock \emph{Proceedings of the National Academy of Sciences}, 114\penalty0
  (30):\penalty0 7908--7914, 2017.

\bibitem[Powell et~al.(2017)Powell, Isler, and Barton]{powell2017}
L.~E. Powell, K.~Isler, and R.~A. Barton.
\newblock Re-evaluating the link between brain size and behavioural ecology in
  primates.
\newblock \emph{Proceedings of the Royal Society B: Biological Sciences},
  284\penalty0 (1865):\penalty0 20171765, 2017.

\bibitem[Dunbar and Shultz(2017)]{dunbar2017}
R.~I.~M. Dunbar and S.~Shultz.
\newblock Why are there so many explanations for primate brain evolution?
\newblock \emph{Philosophical Transactions of the Royal Society B: Biological
  Sciences}, 372\penalty0 (1727):\penalty0 20160244, 2017.

\bibitem[Shultz and Dunbar(2022)]{shultz2022}
S.~Shultz and R.~I.~M. Dunbar.
\newblock Socioecological complexity in primate groups and its cognitive
  correlates.
\newblock \emph{Philosophical Transactions of the Royal Society B},
  377\penalty0 (1860):\penalty0 20210296, 2022.

\bibitem[Lehmann and Dunbar(2009)]{Lehmann2009ProcRSocB}
J.~Lehmann and R.~I.~M. Dunbar.
\newblock Network cohesion, group size and neocortex size in female-bonded old
  world primates.
\newblock \emph{Proceedings of the Royal Society B: Biological Sciences},
  276\penalty0 (1677):\penalty0 4417--4422, 2009.

\bibitem[Pasquaretta et~al.(2014)Pasquaretta, Lev{\'e}, Claidiere, Van~de Waal,
  Whiten, MacIntosh, Pel{\'e}, Bergstrom, Borgeaud, Brosnan, Crofoot, Fedigan,
  Fichtel, Hopper, Mareno, Petit, Schnoell, Polizzi~di Sorrentino, Theirry,
  Tiddi, and Sueur]{Pasquaretta2014SciRep}
C.~Pasquaretta, M.~Lev{\'e}, N.~Claidiere, E.~Van~de Waal, A.~Whiten, A.~J.~J.
  MacIntosh, M.~Pel{\'e}, M.~L. Bergstrom, C.~Borgeaud, S.~F. Brosnan, M.~C.
  Crofoot, L.~M. Fedigan, C.~Fichtel, L.~M. Hopper, M.~C. Mareno, O.~Petit,
  A.~V. Schnoell, E.~Polizzi~di Sorrentino, B.~Theirry, B.~Tiddi, and C.~Sueur.
\newblock Social networks in primates: smart and tolerant species have more
  efficient networks.
\newblock \emph{Scientific Reports}, 4\penalty0 (1):\penalty0 7600, 2014.

\bibitem[Pinter-Wollman et~al.(2014)Pinter-Wollman, Hobson, Smith, Edelman,
  Shizuka, De~Silva, Waters, Prager, Sasaki, Wittemyer, Fewell, and
  McDonald]{Pinterwollman2014BehavEcol}
N.~Pinter-Wollman, E.~A. Hobson, J.~E. Smith, A.~J. Edelman, D.~Shizuka,
  S.~De~Silva, J.~S. Waters, S.~D. Prager, T.~Sasaki, G.~Wittemyer, J.~Fewell,
  and D.~B. McDonald.
\newblock The dynamics of animal social networks: analytical, conceptual, and
  theoretical advances.
\newblock \emph{Behavioral Ecology}, 25\penalty0 (2):\penalty0 242--255, 2014.

\bibitem[Kruase et~al.(2015)Kruase, James, Franks, and Croft]{Krause2015book}
J.~Kruase, R.~James, D.~W. Franks, and D.~P. Croft, editors.
\newblock \emph{Animal Social Networks}.
\newblock Oxford University Press, Oxford, UK, 2015.

\bibitem[Brask et~al.(2021)Brask, Ellis, and Croft]{Brask2021JCompNetw}
J.~B. Brask, S.~Ellis, and D.~P. Croft.
\newblock Animal social networks: an introduction for complex systems
  scientists.
\newblock \emph{Journal of Complex Networks}, 9\penalty0 (2):\penalty0 cnab001,
  2021.

\bibitem[Bansal et~al.(2007)Bansal, Grenfell, and Meyers]{bansal2007}
S.~Bansal, B.~T. Grenfell, and L.~A. Meyers.
\newblock When individual behaviour matters: homogeneous and network models in
  epidemiology.
\newblock \emph{Journal of the Royal Society Interface}, 4\penalty0
  (16):\penalty0 879--891, 2007.

\bibitem[Smith(1982)]{Maynardsmith1982book}
J.~M. Smith.
\newblock \emph{Evolution and the Theory of Games}.
\newblock Cambridge University Press, 1982.

\bibitem[Kappeler and Van~Schaik(2006)]{kappeler2006}
P.~M. Kappeler and C.~P. Van~Schaik.
\newblock \emph{Cooperation in Primates and Humans}.
\newblock Springer, Berlin, Germany, 2006.

\bibitem[No{\"e}(2006)]{noe2006}
R.~No{\"e}.
\newblock Cooperation experiments: coordination through communication versus
  acting apart together.
\newblock \emph{Animal Behaviour}, 71\penalty0 (1):\penalty0 1--18, 2006.

\bibitem[Cheney(2011)]{cheney2011}
D.~L. Cheney.
\newblock Extent and limits of cooperation in animals.
\newblock \emph{Proceedings of the National Academy of Sciences of the United
  States of America}, 108\penalty0 (supplement\_2):\penalty0 10902--10909,
  2011.

\bibitem[Croft et~al.(2015)Croft, Edenbrow, and Darden]{croft2015}
D.~P. Croft, M.~Edenbrow, and S.~K. Darden.
\newblock Assortment in social networks and the evolution of cooperation.
\newblock In \emph{Animal Social Networks}, pages 13--23. Oxford University
  Press, Oxford, UK, 2015.

\bibitem[McAuliffe and Thornton(2015)]{mcauliffe2015}
K.~McAuliffe and A.~Thornton.
\newblock The psychology of cooperation in animals: an ecological approach.
\newblock \emph{Journal of Zoology}, 295\penalty0 (1):\penalty0 23--35, 2015.

\bibitem[Gokcekus et~al.(2021)Gokcekus, Cole, Sheldon, and Firth]{gokcekus2021}
S.~Gokcekus, E.~F. Cole, B.~C. Sheldon, and J.~A. Firth.
\newblock Exploring the causes and consequences of cooperative behaviour in
  wild animal populations using a social network approach.
\newblock \emph{Biological Reviews}, 96\penalty0 (5):\penalty0 2355--2372,
  2021.

\bibitem[Clutton-Brock(2009)]{Cluttonbrock2009Nature}
T.~Clutton-Brock.
\newblock Cooperation between non-kin in animal societies.
\newblock \emph{Nature}, 462\penalty0 (7269):\penalty0 51--57, 2009.

\bibitem[Cheney et~al.(2010)Cheney, Moscovice, Heesen, Mundry, and
  Seyfarth]{cheney2010}
D.~L. Cheney, L.~R. Moscovice, M.~Heesen, R.~Mundry, and R.~M. Seyfarth.
\newblock Contingent cooperation between wild female baboons.
\newblock \emph{Proceedings of the National Academy of Sciences of the United
  States of America}, 107\penalty0 (21):\penalty0 9562--9566, 2010.

\bibitem[Fudenberg and Levine(1998)]{Fudenberg1998book}
D.~Fudenberg and D.~K. Levine.
\newblock \emph{The Theory of Learning in Games}.
\newblock MIT Press, Cambridge, MA, 1998.

\bibitem[Nowak(2006{\natexlab{a}})]{Nowak2006Science}
M.~A. Nowak.
\newblock Five rules for the evolution of cooperation.
\newblock \emph{Science}, 314:\penalty0 1560--1563, 2006{\natexlab{a}}.

\bibitem[Smith(2010)]{smith2010}
E.~A. Smith.
\newblock Communication and collective action: language and the evolution of
  human cooperation.
\newblock \emph{Evolution and Human Behavior}, 31\penalty0 (4):\penalty0
  231--245, 2010.

\bibitem[Salahshour(2020)]{salahshour2020}
M.~Salahshour.
\newblock Coevolution of cooperation and language.
\newblock \emph{Physical Review E}, 102\penalty0 (4):\penalty0 042409, 2020.

\bibitem[{Szab\'{o}} and {F\'{a}th}(2007)]{Szabo2007PhysRep}
G.~{Szab\'{o}} and G.~{F\'{a}th}.
\newblock Evolutionary games on graphs.
\newblock \emph{Physics Reports}, 446:\penalty0 97--216, 2007.

\bibitem[Perc et~al.(2013)Perc, G{\'o}mez-Garde{\~n}es, Szolnoki, Flor{\'\i}a,
  and Moreno]{perc2013}
M.~Perc, J.~G{\'o}mez-Garde{\~n}es, A.~Szolnoki, L.~M. Flor{\'\i}a, and
  Y.~Moreno.
\newblock Evolutionary dynamics of group interactions on structured
  populations: a review.
\newblock \emph{Journal of the Royal Society Interface}, 10\penalty0
  (80):\penalty0 20120997, 2013.

\bibitem[Perc et~al.(2017)Perc, Jordan, Rand, Wang, Boccaletti, and
  Szolnoki]{perc2017}
M.~Perc, J.~J. Jordan, D.~G. Rand, Z.~Wang, S.~Boccaletti, and A.~Szolnoki.
\newblock Statistical physics of human cooperation.
\newblock \emph{Physics Reports}, 687:\penalty0 1--51, 2017.

\bibitem[Tak{\'a}cs et~al.(2021)Tak{\'a}cs, Gross, Testori, Letina, Kenny,
  Power, and Wittek]{takacs2021}
K.~Tak{\'a}cs, J.~Gross, M.~Testori, S.~Letina, A.~R. Kenny, E.~A. Power, and
  R.~P.~M. Wittek.
\newblock Networks of reliable reputations and cooperation: a review.
\newblock \emph{Philosophical Transactions of the Royal Society B: Biological
  Sciences}, 376\penalty0 (1838):\penalty0 20200297, 2021.

\bibitem[Ohtsuki et~al.(2006)Ohtsuki, Hauert, Lieberman, and
  Nowak]{Ohtsuki2006Nature}
H.~Ohtsuki, Ch. Hauert, E.~Lieberman, and M.~A. Nowak.
\newblock A simple rule for the evolution of cooperation on graphs and social
  networks.
\newblock \emph{Nature}, 441\penalty0 (7092):\penalty0 502--505, 2006.

\bibitem[Allen et~al.(2017)Allen, Lippner, Chen, Fotouhi, Momeni, Yau, and
  Nowak]{Allen2017Nature}
B.~Allen, G.~Lippner, Y.-T. Chen, B.~Fotouhi, N.~Momeni, S.-T. Yau, and M.~A.
  Nowak.
\newblock Evolutionary dynamics on any population structure.
\newblock \emph{Nature}, 544\penalty0 (7649):\penalty0 227--230, 2017.

\bibitem[Santos and Pacheco(2005)]{santos2005}
F.~C. Santos and J.~M. Pacheco.
\newblock Scale-free networks provide a unifying framework for the emergence of
  cooperation.
\newblock \emph{Physical Review Letters}, 95\penalty0 (9):\penalty0 098104,
  2005.

\bibitem[Santos et~al.(2006)Santos, Pacheco, and Lenaerts]{santos2006}
F.~C. Santos, J.~M. Pacheco, and T.~Lenaerts.
\newblock Evolutionary dynamics of social dilemmas in structured heterogeneous
  populations.
\newblock \emph{Proceedings of the National Academy of Sciences of the United
  States of America}, 103\penalty0 (9):\penalty0 3490--3494, 2006.

\bibitem[Nowak and May(1992)]{NowakMay1992Nature}
M.~A. Nowak and R.~M. May.
\newblock Evolutionary games and spatial chaos.
\newblock \emph{Nature}, 359\penalty0 (6398):\penalty0 826--829, 1992.

\bibitem[Hauert(2001)]{hauert2001}
Ch. Hauert.
\newblock Fundamental clusters in spatial 2 $\times$ 2 games.
\newblock \emph{Proceedings of the Royal Society B: Biological Sciences},
  268\penalty0 (1468):\penalty0 761--769, 2001.

\bibitem[Aldous and Fill(2002)]{aldous1995}
D.~Aldous and J.~A. Fill.
\newblock Reversible markov chains and random walks on graphs, 2002.
\newblock Unfinished monograph, recompiled 2014 (2002) available at\\
  http://www.stat.berkeley.edu/$\sim$aldous/RWG/book.html. Accessed on August
  6, 2022.

\bibitem[Masuda et~al.(2017)Masuda, Porter, and Lambiotte]{masuda2017}
N.~Masuda, M.~A. Porter, and R.~Lambiotte.
\newblock Random walks and diffusion on networks.
\newblock \emph{Physics Reports}, 716:\penalty0 1--58, 2017.

\bibitem[Nowak et~al.(2004)Nowak, Sasaki, Taylor, and
  Fudenberg]{Nowak2004Nature}
M.~A. Nowak, A.~Sasaki, C.~Taylor, and D.~Fudenberg.
\newblock Emergence of cooperation and evolutionary stability in finite
  populations.
\newblock \emph{Nature}, 428\penalty0 (6983):\penalty0 646--650, 2004.

\bibitem[Nowak(2006{\natexlab{b}})]{Nowak2006book}
M.~A. Nowak.
\newblock \emph{Evolutionary Dynamics}.
\newblock Belknap Press of Harvard University Press, Cambridge, MA,
  2006{\natexlab{b}}.

\bibitem[Sah et~al.(2019)Sah, M{\'e}ndez, and Bansal]{sah2019}
P.~Sah, J.~D. M{\'e}ndez, and S.~Bansal.
\newblock A multi-species repository of social networks.
\newblock \emph{Scientific Data}, 6:\penalty0 44, 2019.

\bibitem[Collier et~al.(2021)Collier, Ali, and Bansal]{collier2021}
M.~Collier, S.~Ali, and S.~Bansal.
\newblock Animal social network repository, August 2021.
\newblock URL \url{https://doi.org/10.5281/zenodo.7595404}.
\newblock Accessed Oct 31, 2022.

\bibitem[Su et~al.(2022)Su, McAvoy, and Plotkin]{Su2022SciAdv}
Q.~Su, A.~McAvoy, and J.~B. Plotkin.
\newblock Evolution of cooperation with contextualized behavior.
\newblock \emph{Science Advances}, 8\penalty0 (6):\penalty0 eabm6066, 2022.

\bibitem[Smaers et~al.(2021)Smaers, Rothman, Hudson, Balanoff, Beatty,
  Dechmann, de~Vries, Dunn, Fleagle, Gilbert, Goswami, Iwaniuk, Jungers,
  Kerney, Ksepka, Manger, Mongle, Rohlf, Smith, Soligo, Weisbecker, and
  Safi]{smaers2021}
J.~B. Smaers, R.~S. Rothman, D.~R. Hudson, A.~M. Balanoff, B.~Beatty, D.~K.~N.
  Dechmann, D.~de~Vries, J.~C. Dunn, J.~G. Fleagle, C.~C. Gilbert, A.~Goswami,
  A.~N. Iwaniuk, W.~L. Jungers, M.~Kerney, D.~T. Ksepka, P.~R. Manger, C.~S.
  Mongle, F.~J. Rohlf, N.~A. Smith, C.~Soligo, V.~Weisbecker, and K.~Safi.
\newblock The evolution of mammalian brain size.
\newblock \emph{Science Advances}, 7\penalty0 (18):\penalty0 eabe2101, 2021.

\bibitem[Newman et~al.(2004)Newman, Jolly, and Rogers]{newman2004}
T.~K. Newman, C.~J. Jolly, and J.~Rogers.
\newblock Mitochondrial phylogeny and systematics of baboons (\emph{Papio}).
\newblock \emph{American Journal of Physical Anthropology}, 124\penalty0
  (1):\penalty0 17--27, 2004.

\bibitem[Smaers et~al.(2019)Smaers, Mongle, Safi, and Dechmann]{smaers2019}
J.~B. Smaers, C.~S. Mongle, K.~Safi, and D.~K.~N. Dechmann.
\newblock Allometry, evolution and development of neocortex size in mammals.
\newblock \emph{Progress in Brain Research}, 250:\penalty0 83--107, 2019.

\bibitem[{R Core Team}(2022)]{R}
{R Core Team}.
\newblock \emph{R: A Language and Environment for Statistical Computing}.
\newblock R Foundation for Statistical Computing, Vienna, Austria, 2022.
\newblock URL \url{https://www.R-project.org/}.

\bibitem[Bartoń(2022)]{barton2022}
K.~Bartoń.
\newblock \emph{MuMIn: Multi-Model Inference}, 2022.
\newblock URL \url{https://CRAN.R-project.org/package=MuMIn}.
\newblock R package version 1.47.1.

\bibitem[Wasserman and Faust(1994)]{wasserman1994book}
S.~Wasserman and K.~Faust.
\newblock \emph{Social Network Analysis: Methods and Applications}.
\newblock Cambridge University Press, New York, NY, 1994.

\bibitem[Newman(2018)]{newman2018book}
M.~Newman.
\newblock \emph{Networks}.
\newblock Oxford University Press, Oxford, UK, 2nd edition, 2018.

\bibitem[Fagiolo(2007)]{fagiolo2007}
G.~Fagiolo.
\newblock Clustering in complex directed networks.
\newblock \emph{Physical Review E}, 76\penalty0 (2):\penalty0 026107, 2007.

\bibitem[Faraway(2016)]{faraway2016}
J.~J. Faraway.
\newblock \emph{Extending the Linear Model with R: Generalized Linear, Mixed
  Effects and Nonparametric Regression Models}.
\newblock Chapman and Hall/CRC, Boca Raton, FL, 2nd edition, 2016.

\bibitem[Burnham and Anderson(2002)]{burnham2002}
K.~P. Burnham and D.~R. Anderson.
\newblock \emph{Model Selection and Multi-Model Inference}.
\newblock Springer-Verlag, New York, NY, 2nd edition, 2002.

\bibitem[Voelkl and Kasper(2009)]{Voelkl2009BiolLett}
B.~Voelkl and C.~Kasper.
\newblock Social structure of primate interaction networks facilitates the
  emergence of cooperation.
\newblock \emph{Biology Letters}, 5\penalty0 (4):\penalty0 462--464, 2009.

\bibitem[Dunbar and Shultz(2021)]{dunbar2021}
R.~I.~M. Dunbar and S.~Shultz.
\newblock Social complexity and the fractal structure of group size in primate
  social evolution.
\newblock \emph{Biological Reviews}, 96\penalty0 (5):\penalty0 1889--1906,
  2021.

\bibitem[Kamilar and Cooper(2013)]{kamilar2013}
J.~M. Kamilar and N.~Cooper.
\newblock Phylogenetic signal in primate behaviour, ecology and life history.
\newblock \emph{Philosophical Transactions of the Royal Society B: Biological
  Sciences}, 368\penalty0 (1618):\penalty0 20120341, 2013.

\bibitem[Collier et~al.(2022)Collier, Albery, McDonald, and
  Bansal]{collier2022}
M.~Collier, G.~F. Albery, G.~C. McDonald, and S.~Bansal.
\newblock Pathogen transmission modes determine contact network structure,
  altering other pathogen characteristics.
\newblock \emph{Proceedings of the Royal Society B: Biological Sciences},
  289\penalty0 (1989):\penalty0 20221389, 2022.

\bibitem[Stephan et~al.(1981)Stephan, Frahm, and Baron]{stephan1981}
H.~Stephan, H.~Frahm, and G.~Baron.
\newblock New and revised data on volumes of brain structures in insectivores
  and primates.
\newblock \emph{Folia Primatologica}, 35\penalty0 (1):\penalty0 1--29, 1981.

\bibitem[Matuschek et~al.(2017)Matuschek, Kliegl, Vasishth, Baayen, and
  Bates]{matuschek2017}
H.~Matuschek, R.~Kliegl, S.~Vasishth, H.~Baayen, and D.~Bates.
\newblock Balancing type i error and power in linear mixed models.
\newblock \emph{Journal of Memory and Language}, 94:\penalty0 305--315, 2017.

\bibitem[Barr et~al.(2013)Barr, Levy, Scheepers, and Tily]{barr2013}
D.~J. Barr, R.~Levy, C.~Scheepers, and H.~J. Tily.
\newblock Random effects structure for confirmatory hypothesis testing: Keep it
  maximal.
\newblock \emph{Journal of Memory and Language}, 68\penalty0 (3):\penalty0
  255--278, 2013.

\bibitem[McElreath(2016)]{mcelreath2016}
R.~McElreath.
\newblock \emph{Statistical rethinking: A {Bayesian} course with examples in
  {R} and {Stan}}.
\newblock Chapman and Hall/CRC, 2016.

\bibitem[Yarkoni(2022)]{yarkoni2022}
T.~Yarkoni.
\newblock The generalizability crisis.
\newblock \emph{Behavioral and Brain Sciences}, 45:\penalty0 1--78, 2022.

\bibitem[Jusup et~al.(2022)Jusup, Holme, Kanazawa, Takayasu, Romi{\'c}, Wang,
  Ge{\v{c}}ek, Lipi{\'c}, Podobnik, Wang, Luo, Klanj{\v{s}}{\v{c}}ek, Fan, and
  Boccaletti]{jusup2022}
M.~Jusup, P.~Holme, K.~Kanazawa, M.~Takayasu, I.~Romi{\'c}, Z.~Wang,
  S.~Ge{\v{c}}ek, T.~Lipi{\'c}, B.~Podobnik, L.~Wang, W.~Luo,
  T.~Klanj{\v{s}}{\v{c}}ek, J.~Fan, and S.~Boccaletti.
\newblock Social physics.
\newblock \emph{Physics Reports}, 948:\penalty0 1--148, 2022.

\bibitem[Ak{\c{c}}ay(2020)]{akcay2020}
E.~Ak{\c{c}}ay.
\newblock Deconstructing evolutionary game theory: coevolution of social
  behaviors with their evolutionary setting.
\newblock \emph{The American Naturalist}, 195\penalty0 (2):\penalty0 315--330,
  2020.

\bibitem[Wu et~al.(2014)Wu, Rong, and Yang]{wu2014}
Z.-X. Wu, Z.~Rong, and H.-X. Yang.
\newblock Community structure benefits the fixation of cooperation under strong
  selection.
\newblock \emph{arXiv preprint arXiv:1408.3267}, 2014.

\bibitem[Alexander(1974)]{alexander1974}
R.~D. Alexander.
\newblock The evolution of social behavior.
\newblock \emph{Annual Review of Ecology and Systematics}, 5:\penalty0
  325--383, 1974.

\bibitem[Freeland(1976)]{freeland1976}
W.~J. Freeland.
\newblock Pathogens and the evolution of primate sociality.
\newblock \emph{Biotropica}, 8\penalty0 (1):\penalty0 12--24, 1976.

\bibitem[Cohen and Havlin(2010)]{cohen2010}
R.~Cohen and S.~Havlin.
\newblock \emph{Complex Networks: Structure, Robustness and Function}.
\newblock Cambridge University Press, Cambridge, UK, 2010.

\bibitem[Krause et~al.(2013)Krause, Krause, Arlinghaus, Psorakis, Roberts, and
  Rutz]{krause2013}
J.~Krause, S.~Krause, R.~Arlinghaus, I.~Psorakis, S.~Roberts, and C.~Rutz.
\newblock Reality mining of animal social systems.
\newblock \emph{Trends in Ecology \& Evolution}, 28\penalty0 (9):\penalty0
  541--551, 2013.

\end{thebibliography}

\end{document}